\documentclass[letterpaper, conference,10pt] {IEEEtran}
\usepackage{amsmath}
\usepackage{cite}
\usepackage{amssymb,amsthm}
\usepackage{bm}
\usepackage{color}
\usepackage{graphicx}
\usepackage{empheq}
\usepackage[linesnumbered,ruled,lined]{algorithm2e}
\usepackage[outdir=./]{epstopdf}
\usepackage{caption}
\usepackage{lipsum}
\usepackage{cite}
\usepackage{multirow}
\usepackage{subcaption}
\usepackage{fancyhdr}
\SetKwInput{Input}{input}
\SetKwInput{Output}{output}

\usepackage{array}
\newcolumntype{L}[1]{>{\raggedright\let\newline\\\arraybackslash\hspace{0pt}}m{#1}}
\newcolumntype{C}[1]{>{\centering\let\newline\\\arraybackslash\hspace{0pt}}m{#1}}
\newcolumntype{R}[1]{>{\raggedleft\let\newline\\\arraybackslash\hspace{0pt}}m{#1}}

\theoremstyle{remark}

\IEEEoverridecommandlockouts
\def\specialpapernotice#1{\if@confmode%
	\def\@specialpapernotice{{\sublargesize\textit{#1}\vspace*{1em}}}%
	\else%
	\def\@specialpapernotice{{\\*[1.5ex]\sublargesize\textit{#1}}\vspace*{-2ex}}%
	\fi}

\begin{document}

	\title{Reconfigurable Intelligent Surface Empowered Rate-Splitting Multiple Access for Simultaneous Wireless Information and Power Transfer}
		\author{	\IEEEauthorblockN{Chengzhong Tian$ ^* $, Yijie Mao$ ^* $, Kangchun Zhao$ ^* $, Yuanming Shi$ ^* $, Bruno Clerckx$ ^{\P\dagger}$ }
		\IEEEauthorblockA{
			$ ^* $School of Information Science and Technology, ShanghaiTech University, Shanghai, China\\
			$ ^\P $Department of Electrical and Electronic Engineering,	Imperial College London, United Kingdom\\
              $^{\dagger}$Silicon Austria Labs (SAL), Graz A-8010, Austria\\
			Email: \{tianchzh, maoyj, zhaokch, shiym\}@shanghaitech.edu.cn, b.clerckx@imperial.ac.uk}
		
	\thanks{This work has been supported in part by the National Nature Science Foundation of China under Grant 62201347; and in part by  Shanghai Sailing Program under Grant 22YF1428400.}

%	\\[-1 ex]
%	{\sublargesize\textit{)}}	
	\\[-3 ex]		
	}
	
\maketitle
\thispagestyle{empty}
\pagestyle{empty}
\begin{abstract}	Rate-splitting multiple access (RSMA) and reconfigurable intelligent surface (RIS) have been both recognized as promising techniques for 6G. The benefits of combining the two techniques to enhance the spectral and energy efficiency have been recently exploited in communication-only networks. Inspired by the recent advances on RIS empowered RSMA, in this work we investigate the use of RIS empowered RSMA for simultaneous wireless information and power transfer (SWIPT) with one transmitter concurrently sending information to multiple  information receivers (IRs) and transferring energy to multiple energy receivers (ERs). Specifically, we jointly optimize the transmit beamformers and the RIS reflection coefficients to maximize the weighted sum-rate (WSR) of IRs under the harvested energy constraint of ERs and the transmit power constraint. An alternating optimization and successive convex approximation (SCA)-based optimization framework is then proposed to solve the problem. Numerical results show that by marrying the benefits of RSMA and RIS, the proposed RIS empowered RSMA achieves a better tradeoff between the WSR of IRs and energy harvested at ERs. Therefore, we conclude that RIS empowered RSMA is a promising strategy for SWIPT.
\end{abstract}


\section{Introduction}

\label{sec:introduction}
\par 
 In recent years, with the rise of the concept of the internet of things (IoT), low-power communications specially catered for IoT nodes with non-replaceable batteries have become indispensable for 6G and beyond. As radio frequency signals carry both information and energy, simultaneous wireless information and power transfer (SWIPT), which concurrently transmits information to the information receivers (IRs) and provides energy for wireless charging for energy receivers (ERs), has become a feasible solution to achieve ubiquitous and self-sustainable wireless communications for future ultra-dense networks with short communication distances \cite{latva2020key}. Meanwhile, as a promising interference management and powerful multiple access technique for 6G, rate-splitting multiple access (RSMA) has significant advantages in improving the system performance in terms of spectral efficiency (SE), energy efficiency (EE), quality of service (QoS), user fairness, robustness against user mobility and channel state information imperfection, etc \cite{9831440}, \cite{7470942}, \cite{clerckx2022primer}. In a practical RSMA model with 1-layer common stream (also known as 1-layer rate-splitting) \cite{7470942}, each user message is split into a common part and a private part. By encoding all common parts into a common stream to be decoded by all receivers and encoding the private parts into private streams respectively for the corresponding users at the transmitter while allowing each user to decode the common and private streams, RSMA achieves to partially decode the interference and partially treat the remaining interference as noise. Such powerful interference management capability makes RSMA more powerful than conventional multiple access techniques such as orthogonal multiple access (OMA), space-division multiple access (SDMA), and power-domain non-orthogonal multiple access (NOMA)\cite{8907421}. RSMA is therefore a promising multiple access for SWIPT. 
\par RSMA empowered SWIPT was first studied in \cite{8815494}. Under a sum energy constraint of ERs, RSMA empowered SWIPT is shown to achieve better WSR performance of IRs than SDMA and NOMA in a wide range of IR and ER deployments. Inspired by its superior performance, RSMA has been studied in SWIPT considering cooperative user relaying \cite{li2021full}, cognitive radio \cite{acosta2020joint}, and imperfect channel state information at the transmitter (CSIT) \cite{camana2022deep}, \cite{su2019robust}. All the aforementioned works have shown the perfect marriage of RSMA and SWIPT. 
To further boost the performance, recent works have initiated the study of applying reconfigurable intelligent surface (RIS) to RSMA empowered SWIPT \cite{camana2022rate}, \cite{pang2022joint}. Specifically, \cite{camana2022rate} investigates the joint transmit beamforming and RIS phase shifts optimization to minimize the transmit power under the QoS rate and energy harvesting constraint at the colocated IRs and ERs with a power-splitting structure. \cite{pang2022joint} extends the single RIS model in \cite{camana2022rate} to double RISs with reflection from one RIS to the other and investigates the user fairness beamforming design. Both \cite{camana2022rate} and \cite{pang2022joint} only investigate a SWIPT model with colocated IRs and ERs. There is a lack of investigation on the RIS empowered RSMA for SWIPT with separated IRs and ERs. 
\par In this work, we propose an RIS empowered RSMA model for SWIPT with separated IRs and ERs. Specifically, the transmit beamforming, common rate allocation, and RIS phase shifts are jointly designed to maximize the WSR of IRs under the harvested energy constraint of the ERs and the transmit power constraint. An alternative optimization-based optimization algorithm is proposed to solve the problem. For the first time, this work illustrates the tradeoff between the WSR of IRs and energy harvested at ERs with or without RIS. For a given ER energy consumption constraint, we show an explicit rate region gain of RIS empowered RSMA over the non-RIS counterparts for SWIPT. Therefore, we draw the conclusion that RIS empowered RSMA is capable of further boosting the spectral efficiency of SWIPT with separated IRs and ERs. It is a promising transmission strategy for 6G.    
\section{System Model and Problem Formulation}
\label{sec: system model}	\par Consider an RIS-assisted RSMA for a multi-antenna SWIPT transmission network consisting of one multi-antenna base station (BS), one RIS with $N$ elements, $K$ single-antenna IRs, and $J$ single-antenna ERs. The BS is equipped with $N_t$ antennas, the IRs and ERs are respectively indexed by $\mathcal{K}=\left\{1,2,\cdots,K\right\}$ and $\mathcal{J}=\left\{1,2,\cdots,J\right\}$. The reflecting elements at the RIS are indexed by $\mathcal{N}=\left\{1,2,\cdots,N\right\}$. As per Fig. 1, the BS simultaneously transmits information to IRs and energy to ERs with the assistance of an RIS. The transmit signal $\mathbf{x}\in\mathbb{C}^{N_t\times1}$ contains both the information signals for IRs and the energy signals for ERs.
\begin{figure}[t!] 
 \vspace{-4mm}
	\centering
	\includegraphics[width=0.45\textwidth]{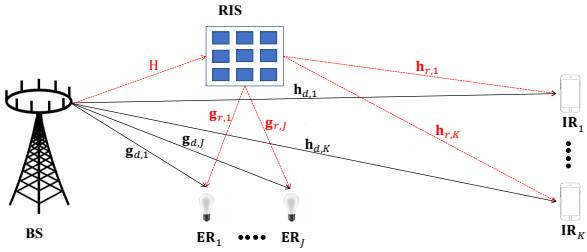}
	\caption{ A multi-antenna multi-user transmission network with RIS empowered SWIPT}
 \vspace{-5mm}
\end{figure}
RSMA is considered for information transmission. Specifically, the message of IR-$k$ is split into a common part $W_{c,k}$ and a private parts $W_{p,k}$, $\forall{k}\in{\mathcal{K}}$. The common parts of all IRs $\left\{W_{c,1}\cdots,W_{c,K}\right\}$ are combined into a common message $W_c$ which is encoded to common stream $s_c^{\text{I}}$, while the private parts $\left\{W_{p,1}\cdots,W_{p,K}\right\}$ are independently encoded to the private streams $\left\{s_1^{\normalfont \text{I}},\cdots,s_K^{\text{I}}\right\}$. The set of streams for IRs ${\bf s}^{\text{I}}=\left[s_c^{\text{I}},s_1^{\text{I}},\cdots,s_K^{\text{I}}\right]^T\in \mathbb{C}^{(K+1)\times1}$ and ERs ${\bf s}^{\text{E}}=\left[s_1^{\text{E}},\cdots,s_J^{\text{E}}\right]^T\in \mathbb{C}^{J\times1}$ are linearly precoded by 
$\mathbf{P}=\left[ {{\mathbf p}_c,{\mathbf p}_1,\cdots,{\mathbf p}_K}\right]$
and $\mathbf{F}=\left[ {\mathbf f}_1,\cdots,{\mathbf f}_J\right]$, where ${{\mathbf p}_c} \in \mathbb{C}^{N_{t}\times1}$, ${{\bf p}_k} \in \mathbb{C}^{N_{t}\times1}$, ${\bf f}_j\in \mathbb{C}^{N_t\times1}$ are the respective precoders for the common stream, private stream for IR-$k$, and the energy signal for ER-$j$. The transmit signal is:
\begin{equation}
	{\bf x}={\bf p}_c s_c^{\text{I}}+\sum_{k\in{\mathcal{K}}}{{\bf p}_k}s_k^{\text{I}}+\sum_{j\in{\mathcal{J}}}{{\bf f}_j}s_j^{{\text{E}}}.\\
\end{equation}
\par  We assume that $\mathbb{E}\left\{{\bf s}^{\text {I}}\left({{\bf s}^{\text {I}}}\right)^H\right\}= \mathbb{E}\left\{{\bf s}^{\text {E}}\left({{\bf s}^{\text {E}}}\right)^H\right\}=\bf I$ and the transmit power is limited to $P_t$. The transmit signal goes through both the direct channel between the BS and the IRs/ERs and the assisted channel from the BS to the RIS to the IRs/ERs. By controlling the reflecting phase shift $\theta_u \in \left[0,2\pi \right]$ for each reflecting unit $u$, the RIS is able to reflect the signal towards IRs and ERs. Let $\boldsymbol \theta=[\theta_1,\cdots,\theta_N]$
and $\boldsymbol {\Theta} = $ diag($e^{j\theta_1},\cdots,e^{j\theta_N}$) be the reflecting phase shift vector and the corresponding phase shift matrix, respectively. ${\bf h}_{d,k}$ and $ {{\bf g}_{d,j}}$ respectively denote the channels between the BS and IR-$k$ as well as the BS and ER-$j$, while ${\bf h}_{r,k}$ and ${\bf g}_{r,j}$ represent the channel between the RIS and IR-$k$ as well as the RIS to ER-$j$, respectively. ${\bf H }\in\mathbb C^{N\times {N_t}} $ denotes the channel matrix from the BS to the RIS. The respective effective channels from the BS to IR-$k$ and ER-$j$ ${{\bf h}_k^H}$ and ${{\bf g}_j^H}$ are given by\\
\begin{equation}
	\begin{aligned}
		{{\bf h}_k^H} = {\bf h}_{r,k}^H{\bf \Theta H} + {\bf h}_{d,k}^H,\forall{k}\in{\mathcal{K}},\\
		{{\bf g}_j^H}=  {\bf g}_{r,j}^H {\bf\Theta H} +{\bf g}_{d,j}^H,\forall{j}\in{\mathcal{J}}.
	\end{aligned}
\end{equation}
\par Let $n_k^{\text{I}}$ and $n_j^{\text {E}}$ be the
respective Additive White Gaussian Noises (AWGNs) received
at IR-$k$ and ER-$j$ with zero mean and variance $\sigma_k^2$ or $\sigma_j^2$, respectively. Perfect CSIT
is assumed at all nodes. The signals received at IR-$k$ and ER-$j$ are respectively described as \begin{equation}
	\begin{aligned}
		y_k^{\text{I}}&={{\bf h}_k^H\bf x}+n_k^{\text{I}},\forall{k}\in{\mathcal{K}},\\
		y_j^{\text{E}}&={{\bf g}_j^H\bf x}+n_j^{\text{E}},\forall{j}\in{\mathcal{J}}.
	\end{aligned}
\end{equation}
The energy signal $s_j^E$ does not carry any information, and is assumed to be perfectly known at the transmitter and IRs. Therefore, IRs are able to eliminate interference from the energy signals before decoding the intended signals. 
\par	The transmission rates of the common stream $s_c^{\text{I}}$ and the private stream $s_k^{\text{I}}$ at IR-$k$ are respectively given as\begin{equation}
	\begin{aligned}
		R_{c,k} &= \log_2\left(1+\frac{\left|{\bf  h}_k^H{\bf p}_c\right|^2}{\sum_{j\in{\mathcal{K}}}\left|{\bf  h}_k^H{\bf p}_j\right|^2+\sigma_k^2}\right),\forall{k}\in{\mathcal{K}},\\R_{k} &= \log_2\left(1+\frac{\left|{\bf  h}_k^H{\bf p}_k\right|^2}{\sum_{j\in{\mathcal{K}},j\neq{k}}\left|{\bf  h}_k^H{\bf p}_j\right|^2+\sigma_k^2}\right),\forall{k}\in{\mathcal{K}}.
	\end{aligned}
\end{equation}
\par To ensure the common message can be decoded successfully by every IR, the achievable rate of the common stream shall not exceed
$R_c= \min \left\{R_{1,c},\cdots,R_{K,c}\right\}$. As $R_c$ is shared
by $K$ IRs, we have $\sum_{k\in{\mathcal {K}}}C_k=R_c$,  
where $C_k$ represents the part of the common rate $R_c$ used to transmit $W_{c,k}$. The total achieve rate of IR-$k$ is given by 
$R_{k,tot}=C_k+R_k$.
\par	Assuming a linear model of the energy harvester\footnote{The linear model of the energy harvester is considered in this paper, the study of the nonlinear model of the energy harvester\cite{8476597} will be studied in the future work.}, the harvested energy $Q_j$ is proportional to the power received at ER-$j$, which is given as
\begin{equation}
	Q_j=\zeta \left(\left|{\bf  g}_j^H{\bf p}_c\right|^2+\sum_{k\in{\mathcal{K}}}\left|{\bf  g}_j^H{\bf p}_k\right|^2+\sum_{j'\in{\mathcal{J}}}\left|{\bf  g}_j^H{\bf f}_{j'}\right|^2\right),\\
	\forall j\in {\mathcal{J}},\\
\end{equation}
where $\zeta$ is the energy conversion efficiency and $0\leq \zeta\leq 1$.
\par In this work, to find the optimal rate and energy tradeoff between the IRs and ERs, we jointly design the precoders $\mathbf{P}, \mathbf{F}$, common rate allocation $\mathbf{c}=[C_1,\cdots, C_K]$ and RIS phase shifts $\bm{\theta}$ to maximize the WSR of IRs for a certain sum harvested energy lower bound $E_{th}$ at the ERs and the transmit power consumption upper bound $P_t$ at the BS. Let $u_k$ denote the weight allocated to IR-$k$, the problem for the proposed RIS empowered RSMA model for SWIPT is formulated as
\begin{subequations}
	\begin{align}
		\underset{\bf P,F,c,{\boldsymbol{\theta}}}{\max}&\sum_{k\in{\mathcal{K}}}u_k(C_k+R_k)\\
	%\end{align}
%	\begin{align}
		\operatorname{ s.t.}   &{\sum_{k\in{\mathcal{K}}}C_k\leq R_{c,k}},\forall{k}\in{\mathcal{K}},\\
		&{\sum_{j\in{\mathcal{J}}}Q_j\geq E_{th}},\\
		&\text {tr}(\textbf{PP}^H)+\text {tr}(\textbf{FF}^H)\leq P_t,\\
		&\textbf{c}\geq \bf 0,\\
	%\end{align}\begin{align}
		&0\leq\theta_n\leq 2\pi,\forall{n}\in{\mathcal{N}}.
	\end{align}
\end{subequations}
Constraint (6b) ensures the successful decoding of the common stream at all IRs. (6c) is the sum harvested energy constraint at the ERs and (6d) is the transmit power constraint at the BS. In order to solve the non-convexity of problem (6), we propose an optimization framework based on alternative optimization, which is delineated in the next section.
\section{Proposed Optimization Framework}
\par To solve problem (10), we first decompose the problem into two subproblems of beamforming $\{\mathbf{P}, \mathbf{F}, \mathbf{c}\}$ and RIS phase shifts $\{\bm{\theta}, \mathbf{c}\}$, which are solved in an iterative manner. Since the user effective channels are reconfigured after tuning the RIS phase shifts or the beamformers, the common rate is changed accordingly and the common rate allocation is therefore required to be optimized in both subproblems. In the following two subsections, the optimization approaches to solve the two subproblems are specified followed by an overall optimization framework.
\subsection{Beamforming and Common Rate Optimization}
\par For a given RIS phase shifts $\bm{\theta}$, we follow \cite{8815494} and use a weighted minimum mean squared error (WMMSE) and successive convex approximation (SCA)-based algorithm to find the
near optimal precoders $\mathbf{P}, \mathbf{F}$ and common rate allocation $\mathbf{c}$. The augmented weighted MSEs are
defined as 
$\xi_{c,k}=w_{c,k}\epsilon_{c,k}-\log_2(w_{c,k})$, $\xi_{k}=w_{k}\epsilon_{k}-\log_2(w_{k})$,
where $w_{c,k}$ and $w_k$ are the weights of the MSEs of the common and private streams at IR-$k$. 
The optimal weights could be denoted as $w_{c,k}=w_{c,k}^{\text{MMSE}}\triangleq(\epsilon_{c,k}^{\text{MMSE}})^{-1}$,	$w_{k}=w_{k}^{\text{MMSE}}\triangleq(\epsilon_{k}^{\text{MMSE}})^{-1}$ by solving $\frac{\partial\xi_{c,k}(g_{c,k}^{\text{MMSE}})}{\partial w_{c,k}}=0$ and  $\frac{\partial\xi_{k}(g_{c,k}^{\text{MMSE}})}{\partial w_{k}}=0$.
Thus, the relationship between MSEs and rate can be established as
$ 	\xi_{c,k}^{\text{MMSE}}\triangleq 1-R_{c,k}$ and	$\xi_{k}^{\text{MMSE}}\triangleq 1-R_{k}$.
\par In order to solve the non-convexity of constraint
(6c), we carry out the first-order Taylor expansion to the harvested
energy at each user. Based on the first-order lower bound of
$\left|{\bf g}_j^H{\bf p}_k\right|^2$
at a given point ${\bf p}^{[n]}_k$, which is given by
\begin{equation}
	\begin{split}
		\left|{\bf g}_j^H{\bf p}_k\right|^2\geq2R(({\bf p}_k^{[n]})^H{\bf g}_j{\bf g}_j^H{\bf p}_k)-\left|{\bf g}_j^H{\bf p}_k^{[n]}\right|^2
		\triangleq \Psi^{[n]}({\bf p}_k, {\bf g}_j),
	\end{split}
\end{equation}
constraint (6c) becomes
\begin{equation}
\begin{aligned}
	&\sum_{j\in{\mathcal{J}}}\zeta\left(\Psi^{[n]}({\bf p}_c,{\bf g}_{j})+\sum_{k\in{\mathcal{K}}}\Psi^{[n]}({\bf p}_k,{\bf g}_{j})\right.\\&\left.+\sum_{j'\in{\mathcal{J}}}\Psi^{[n]}({\bf f}_{j'},{\bf g}_{j})\right)\geq E_{th}.
	\end{aligned}
\end{equation}
\par Then, the original
problem (6) is reformulated as
\begin{subequations}
	\begin{align}
		\underset{\bf P,F,x,w,g}{\min}&\sum_{k\in{\mathcal{K}}}u_k(X_k+\xi_k)\\
		\operatorname{ s.t.}   \quad&\sum_{k\in{\mathcal{K}}}X_k+1\geq \xi_{c,k},\forall{k}\in{\mathcal{K}},\\ 
		&\text {tr}({\textbf{PP}}^H)+\text {tr}(\textbf{FF}^H)\leq P_t,\\
	&	\textbf{x}\leq \bf0,\\
	&(8),\nonumber
	\end{align}
\end{subequations}
where $X_k=-C_k$ and ${\bf x} = [X_1,\cdots,X_K]$ is the
WMSE vector. $ {\bf w} = [w_1,\cdots, w_K, w_{1,c},\cdots,w_{K,c}] $ is the vector
of all MSE weights. ${\bf g} = [g_1,\cdots,g_K,g_{1,c},\cdots, g_{K,c}]$ is the
vector containing all equalizers. Ploblem (9) is convex and can be solved by CVX toolbox direction [12]. 
The overall process of the SCA method to solve beamforming problem is illustrated in Algorithm 1, where $\mathrm{WMMSE}=\sum_{k\in{\mathcal{K}}}u_k(X_k+\xi_k)$ and $\mathrm{WSR}=\sum_{k\in{\mathcal{K}}}u_k(C_k+R_k)$ are the calculated WMMSE and WSR, respectively.
\begin{algorithm}
	\caption{WMMSE and SCA-based algorithm}
	{\bf  Initialize:} $m \leftarrow 0$, $n \leftarrow 0$, ${\bf P}^{[n]}$, ${\bf F}^{[n]}$, $\mathrm{WMMSE}^{[n]}$, ${\mathrm{WSR}^{[m]} }$, ${\bf w}$, ${\bf g}$,  ${\bm \theta}$\;
	\Repeat{$\left|\mathrm{WSR}^{[m]} - \mathrm{WSR}^{[m-1]}\right|\leq \epsilon$}{\Repeat{ $\left|\mathrm{WMMSE}^{[n]} - \mathrm{WMMSE}^{[n-1]}\right|\leq \epsilon$}{
	   update (${\bf P}^{[n+1]}$, ${\bf F}^{[n+1]}$, ${\bf x}^{[n+1]}$) by solving 
	 problem (9) using $\bf w$, $\bf g$ and ${\bf P}^{[n]}$, ${\bf F}^{[n]}$;\\
	  update $\normalsize \mathrm{WMMSE}^{[n+1]}$ using (${\bf P}^{[n+1]}$, ${\bf F}^{[n+1]}$);\\
		   ${\bf P}^{*}\leftarrow{\bf P}^{[n+1]}$, ${\bf F}^{*}\leftarrow{\bf F}^{[n+1]}$, $n \leftarrow n + 1$;\\} 	
	  $m \leftarrow m + 1$,	${\bf P}^{[m]}\leftarrow{\bf P}^{*}$, ${\bf F}^{[m]}\leftarrow{\bf F}^{*}$, ${\bf w}\leftarrow{\bf w}^{\mathrm{MMSE}}$. ${\bf g}\leftarrow{\bf g}^{\mathrm{MMSE}}$;\\	
		${\bf c}^{[m]}=-{\bf x}^{[m]}$;\\
		update $\normalsize \mathrm{WSR}^{[m]}$ using (${\bf P}^{[m]}$, ${\bf c}^{[m]}$);}
\end{algorithm}
 \vspace{-2mm}
\subsection{Common rate and Phase Optimization}
\par	For given precoders $\mathbf{P}, \mathbf{F}$ and common rate allocation $\mathbf{c}$, we then optimize RIS phase shifts $\bm{\theta}$ based on the SCA approach. By introducing the slack variable vector $\bm{\eta} = [\eta_1,\cdots,\eta_K]^T$ and slack variable ${\eta_t}$ to denote the SINRs of the private and common streams, and defining ${\bf a}_{ki} \triangleq (\text{diag}({\bf h}^H_{r,k} ){\bf Hp}_i)$, ${\bf a}_{kc} \triangleq (\text{diag}({\bf h}^H_{r,k} ){\bf Hp}_c)$, ${\bf b}_{jj'} \triangleq (\text{diag}({\bf g}^H_{r,j} ){\bf Hf}_{j'})$, $s_n\triangleq e^{j\theta_n}$ and ${\bf s} \triangleq [s_1,\cdots,s_N ]^T$. With the help of ${\bf a}_{ki} $, ${\bf a}_{kc} $, ${\bf b}_{jj'} $ and ${\bf s}$, we can show ${\bf h}_{r,k}^H{\bf \Theta H}{\bf p}_k={\bf a}_{kk}^H\bm{s}$, $ {\bf g}_{r,j}^H {\bf\Theta H}{\bf f}_{j'}={\bf b}_{jj'}^H\bm{s}$, the original problem is equivalently reformulated as 
\begin{subequations}
	\begin{align}
			\underset{{\boldsymbol{\theta}},{\boldsymbol{\eta}},{\boldsymbol{c}},{{\eta_t}}}{\max} &\sum_{k\in{\mathcal{K}}}u_k(C_k+\log_2(1+\eta_k))\\
		\operatorname{ s.t.}  
		&\frac{\left|{\bf a}_{kc}^H{\bf s}+{\bf h}_{d,k}^H{\bf p}_c\right|^2}{\sum_{i\in{\mathcal{K}}}\left| {\bf a}_{ki}^H{\bf s}+{\bf h}_{d,k}^H{\bf p}_i\right|^2+\sigma_k^2}\geq \eta_t,\forall{k}\in{\mathcal{K}},\\
	&\frac{\left|{\bf a}_{kk}^H\bm{s}+{\bf h}_{d,k}^H{\bf p}_k\right|^2}{\sum_{i\in{\mathcal{K}},i\neq{k}}\left|{\bf a}_{ki}^H\bm s+{\bf h}_{d,k}^H{\bf p}_i\right|^2+\sigma_k^2}\geq\eta_k,\forall{k}\in{\mathcal{K}},\\
		&\sum_{j\in{\mathcal{J}}}\zeta \left(\left|{\bf b}^H
		_{jc}{\bf s}+ {\bf g}_{d,j}^H{\bf p}_c\right|^2+\sum_{k\in{\mathcal{K}}}\left|{\bf b}^H
		_{jk}{\bf s}+ {\bf g}_{d,j}^H{\bf p}_k\right|^2\nonumber\right.\\
	&\left.	+\sum_{j'\in{\mathcal{J}}}\left|{\bf b}^H
		_{jj'}{\bf s}+ {\bf g}_{d,j}^H{\bf f}_{j'}\right|^2\right)\geq E_{th},\\
			&\log_2(1+\eta_t)\geq \sum_{k\in\mathcal{K}}C_k,\\
	&	\left|s_n\right|=1,\forall{n}\in{\mathcal{N}}.
	\end{align}
\end{subequations}
\par To deal with the non-convexity of constraint (10d), we
carry out the first-order Taylor approximation to the harvested
energy at each user. Based on the first-order lower bound of
$\left|{\bf b}^H
_{jj'}{\bf s}+ {\bf g}_{d,j}^H{\bf f}_{j'}\right|^2$
at a given point ${\bf s}^{[t-1]}$, which is given by\begin{equation}
\begin{aligned}
&\left|{\bf b}^H_{jj'}{\bf s}+ {\bf g}_{d,j}^H{\bf f}_{j'}\right|^2
	\\&\geq2\mathcal{R}(({\bf b}_{jj'}{\bf s}^{(t-1)}+ {\bf g}_{d,j}^H{\bf f}_{j'})^H{\bf b}_{jj'}{\bf s})-\left|{\bf b}_{jj'}{\bf s}^{(t-1)}+ {\bf g}_{d,j}^H{\bf f}_{j'}\right|^2\\&\triangleq Q'^{(t-1)}({\bf f}_{j'},{\bf g}_{d,j}),
	\end{aligned}
\end{equation}
where $s_n^{(t-1)}$ is the 
value of the variable $s_n$ at the $t-1$ iteration. With the help of (11), constraint (10d) is rewritten as
\begin{equation}\begin{aligned}
&\sum_{j\in{\mathcal{J}}}\zeta\left(Q'^{(t-1)}({\bf p}_c,{\bf g}_{d,j})+\sum_{k\in{\mathcal{K}}}Q'^{(t-1)}({\bf p}_k,{\bf g}_{d,j})\right.\\&\left.+\sum_{j'\in{\mathcal{J}}}Q'^{(t-1)}({\bf f}_{j'},{\bf g}_{d,j})\right)
\geq E_{th}.
\end{aligned}
\end{equation}
\par To handle the non-convexity constraint (10f), we add a compensation term to the objective function and problem (10) is rewritten as
\begin{subequations}
	\begin{align}
		\underset{{\bf s},{\boldsymbol{\eta}},{\boldsymbol{c}},{{\eta_t}}}{\max}&\sum_{k\in{\mathcal{K}}}u_k(C_k+\log_2(1+\eta_k)))
		+C\sum_{k\in{\mathcal{K}}}(\left|s_n\right|^2-1)\\
		\operatorname{ s.t.}  
	&	\left|s_n\right|\leq1 ,\forall{n}\in{\mathcal{N}},\\
		&(10\text{b}),(10\text{c}),(10\text{e}),(12),\nonumber
	\end{align}
\end{subequations}
where $C$ is a positive constant. Constraints (10b), (10c) and (13a) remain non-convex. Following \cite{9145189}, we use the first-order Taylor expansion to find the upper bound of $C\sum_{k\in{\mathcal{K}}}(\left|s_n\right|^2-1)$ at a given points ${\bf s}^{[t-1]}$, which is $C\sum_{k\in{\mathcal{K}}}(\left|s_n\right|^2-1)\leq 2C\sum_{k\in{\mathcal{K}}}s_n^{(t-1)}(s_n-s_n^{(t-1)})$ and approximate (13a) by
	\begin{equation}
	\begin{aligned}
			\underset{{\bf s},{\boldsymbol{\eta}},{\boldsymbol{c}},{{\eta_t}}}{\max} \sum_{k\in{\mathcal{K}}}&u_k(C_k+\log_2{(1+\eta_k)})\\&+2C\sum_{k\in{\mathcal{K}}}s_n^{(t-1)}(s_n-s_n^{(t-1)})
	\end{aligned}
	\end{equation}
\par To handle the
non-convexity of constraints (10b) and (10c), we introduce variables $\beta_1,\cdots,\beta_K,\beta_t$ and
constraints (10b) and (10c) are equivalent to:
\begin{equation}\begin{aligned}
	\left|{\bf a}_{kk}{\bf s}+ {\bf h}_{d,k}^H{\bf p}_k\right|^2&\geq\beta_k\eta_k,\forall{k}\in{\mathcal{K}},\\&=\frac{1}{4}((\beta_k+\eta_k)^2-(\beta_k-\eta_k)^2),
	\end{aligned}
\end{equation}
\begin{equation}\begin{aligned}
	\left|{\bf a}_{kc}{\bf s}+ {\bf h}_{d,k}^H{\bf p}_c\right|^2&\geq\beta_{t}\eta_{t},\forall{k}\in{\mathcal{K}},\\&=\frac{1}{4}((\beta_{t}+\eta_{t})^2-(\beta_{t}-\eta_{t})^2),
	\end{aligned}
\end{equation}
\begin{equation}
	\sum_{i\in{\mathcal{K}},i\neq k}\left|{\bf a}_{ki}{\bf s}+ {\bf h}_{d,k}^H{\bf p}_i\right|^2+\sigma_k^2\leq\beta_k,\forall{k}\in{\mathcal{K}},
\end{equation}
\begin{equation}
	\sum_{i\in{\mathcal{K}}}\left|{\bf a}_{ki}{\bf s}+ {\bf h}_{d,k}^H{\bf p}_i\right|^2+\sigma_k^2\leq\beta_{t},\forall{k}\in{\mathcal{K}}.
\end{equation}
\par By further using the first-order Taylor expansion at the given point $\{{\bf s}^{(t-1)},\beta_{k}^{(t-1)},\eta_{k}^{(t-1)},\beta_{t}^{(t-1)},\eta_{t}^{(t-1)}\}$, constraints (15), (16) can be approximated by
\begin{equation}\begin{aligned}
		&2\mathcal{R}(({\bf a}_{kk}{\bf s}^{(t-1)}+ {\bf h}_{d,k}^H{\bf p}_k)^H{\bf a}_{kk}{\bf s})-\left|{\bf a}_{kk}{\bf s}^{(t-1)}\right|^2+ \left|{\bf h}_{d,k}^H{\bf p}_k\right|^2\\
		\geq&\frac{1}{4}\left((\beta_k+\eta_k)^2-2(\beta_k^{(t-1)}-\eta_k^{(t-1)})(\beta_k-\eta_k)\right.\\
		&\left.+(\beta_k^{(t-1)}-\eta_k^{(t-1)})^2\right),\forall{k}\in{\mathcal{K}},
	\end{aligned}
\end{equation}
\begin{equation}
\begin{aligned}
		&2\mathcal{R}(({\bf a}_{kc}{\bf s}^{(t-1)}+ {\bf h}_{d,k}^H{\bf p}_c)^H{\bf a}_{kc}{\bf s})-\left|{\bf a}_{kc}{\bf s}^{(t-1)}\right|^2+ \left|{\bf h}_{d,k}^H{\bf p}_c\right|^2\\
		\geq&\frac{1}{4}\left((\beta_t+\eta_{t})^2-2(\beta_t^{(t-1)}-\eta_{t}^{(t-1)})(\beta_t-\eta_{t})\right.\\
		&\left.+(\beta_t^{(t-1)}-\eta_{t}^{(t-1)})^2\right),\forall{k}\in{\mathcal{K}},
	\end{aligned}
\end{equation}
 where the left-hand side of inequality is the first-order lower bound of $\left|{\bf a}_{kk}{\bf s}+ {\bf h}_{d,k}^H{\bf p}_k\right|^2$, the right-hand side of inequality are the first-order upper bound of $\frac{1}{4}((\beta_k+\eta_k)^2-(\beta_k-\eta_k)^2)$ and $\frac{1}{4}((\beta_{t}+\eta_{t})^2-(\beta_{t}-\eta_{t})^2)$. With the above approximations, problem (13) can be approximated by the following convex
problem:
\begin{equation}
	\begin{aligned}
			\underset{{\bf s},{\boldsymbol{\eta}},{\boldsymbol{\beta}},{\boldsymbol{c}},{\eta_t}}{\max} &\sum_{k\in{\mathcal{K}}}u_k(C_k+\log_2{(1+\eta_k)})\\&+2C\sum_{k\in{\mathcal{K}}}s_n^{(t-1)}(s_n-s_n^{(t-1)})\\
		\operatorname{s.t.}\quad	 
		&(10\text{e}),(12),(13\text{b}),(17),(18),(19),(20),
	\end{aligned}
\end{equation}
\begin{algorithm}
	\caption{Common rate and Phase Optimization with SCA}
{\bf	Initialize:} $t \leftarrow 0$, ${\bf s}^{[t]}$, $\boldsymbol\beta^{[t]}$, $\eta_t^{[t]}$, $\boldsymbol\eta^{[t]}$, $\boldsymbol c^{[t]}$, $\mathrm{OBJ}^{[t]}$; \\ 
\Repeat{$\left|\mathrm{OBJ}^{[t]} - \mathrm{OBJ}^{[t-1]}\right|\leq \epsilon$}{ update (${\bf s}^{[t+1]}$, $\boldsymbol\beta^{[t+1]}$, $\boldsymbol\eta^{[t+1]}$, $\eta_t^{[t+1]}$, $\boldsymbol c^{[t+1]}$) by solving problem (21) using ${\bf s}^{[t]}$, $\boldsymbol\beta^{[t]}$, $\boldsymbol\eta^{[t]}$ and 
	the given ${\bf P}^{*}$, ${\bf F}^{*}$; \\
	update $\mathrm{OBJ}^{[t+1]}$ using $\boldsymbol\eta^{[t+1]}$ and 
	the  ${\bf c}^{[t+1]}$; \\
		$t\leftarrow t+1$;} 
\end{algorithm}
where ${\bm \beta}=[\beta_1,\cdots, \beta_K]^T$. Problem (21) can be solved by the standard CVX toolbox. The
detailed process of using the SCA to solve problem
(21) is illustrated in Algorithm 2, where $\mathrm{OBJ}=\sum_{k\in{\mathcal{K}}}u_k(C_k+\log_2(1+\eta_k)$ is the calculated objective function in each iteration.
\subsection{Alternative Optimization Algorithm}
The overall AO algorithm to jointly optimize precoders ${\bf{P}}$, ${\bf{F}}$, the common rate ${\bf{c}}$ and RIS phase shifts ${\boldsymbol{\theta}}$ is illustrated in Algorithm 3. Specifically, the optimal precoders ${\bf{P}}$, ${\bf{F}}$ and common rate $ {\bf{c}}$ are updated with the fixed phase by Algorithm 1. After that, the optimal phase ${\boldsymbol{\theta}}$ and common rate ${\bf{c}}$ are jointly updated with the fixed precoders ${\bf{P}}$ and ${\bf{F}}$ by Algorithm 2. This process continues until convergence.

\begin{algorithm}
\caption{AO algorithm}
{\bf  Initialize:} $p \leftarrow 0$, ${\bm \theta^{[p]}}$, ${\bf P}^{[p]}$, ${\bf F}^{[p]}$, ${\bf c}^{[p]}$, ${ \mathrm{WSR}^{[p]}}$; \\
\Repeat{$\left|\mathrm{WSR}^{[p]} - \mathrm{WSR}^{[p-1]}\right|\leq \epsilon$}{	$p \leftarrow p + 1$;\\
   update (${\bf P}^{[p]}$, ${\bf F}^{[p]}$, ${\bf c}^{[p]}$) by using {\bf Algorithm 1} with fixed ${\bm \theta}$;\\

	 update (${\bm \theta^{[p]}}$, ${ \bf c}^{[p]}$) by using {\bf Algorithm 2} with fixed ${\bf P}^{[p]}$, ${\bf F}^{[p]}$;\\
	 
	 update $\mathrm{WSR}^{[p]}$ with ${\bf P}^{[p]}$, ${\bf F}^{[p]}$, ${\bm \theta^{[p]}}$, ${ \bf c}^{[p]}$;\\}
\end{algorithm}
\textit{Convergence Analysis:}  
The proposed Algorithm 3 contains two loops. In the inner loop, Algorthm 1 and Algorithm 2 are conducted. Algorithm 1 and Algorithm 2 are conducted.  guarantees the WSR is monotonically increasing, and the solution at iteration $m$  is a feasible solution of problem (9). Moreover, constraint (9c) ensures that the existence of an upper bound on WSR. Therefore, the convergence of Algorithm 1 is guaranteed. The solution at iteration $t$ by Algorithm 2 is also a feasible solution of problem (21). Furthermore, 
the SCA Algorithm ensures that the objective function is monotonically increasing. Due to the RIS phase shift constraint (13b), the objective function has an upper bound, thus ensuring the convergence of Algorithm 2. 
In the outer loop, the solution of the beamforming problem is a feasible solution of the RIS phase shift problem, vise versa. Furthermore, WSR has an upper bound due to constraints (6c) and (13b), which ensures the convergence of Algorithm 3.
\section{Numerical Results}
In this section, We illustrate the performance of RIS empowered RSMA for SWIPT and compare it with the following schemes:
\begin{figure}[t!]	
	\centering
	\includegraphics[width=0.43\textwidth]{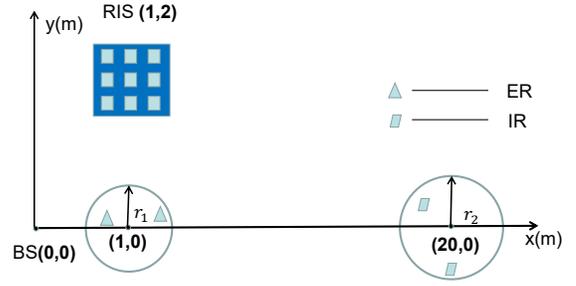}
	\caption{The location distribution of BS, RIS, IRs
and ERs}
	\label{fig:The location distribution of BS, RIS, IRs
and ERs}
 \vspace{-4mm}
\end{figure}
\begin{enumerate}
 \item \textbf{RSMA+RIS}—This is the scheme proposed in Section II
\par \item \textbf{RSMA}—This is the case without RIS in case 1) as studied in \cite{8815494}.
\par \item \textbf{SDMA+RIS}—This is a special case when common rate $R_c$ in case 1) is 0.
\par \item \textbf{SDMA}—This is the case without RIS in case 3).
\par \item \textbf{NOMA+RIS}—This is the case that RIS empowered NOMA for SWIPT.
\par \item \textbf{NOMA}—This is the case without RIS in case 5).
\end{enumerate}
\begin{figure}[t!]	
	\centering
	\includegraphics[width=0.43\textwidth]{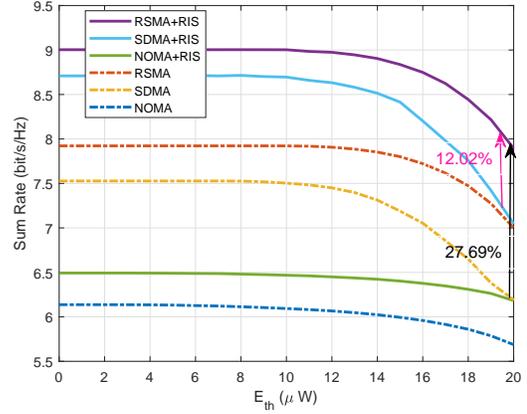}
	\caption{Rate-energy region comparison of different strategies with $P_t=10 \text{dB}$.}
 \vspace{-4mm}	
\end{figure}
\begin{figure}[t!]
	\centering
	\includegraphics[width=0.43\textwidth]{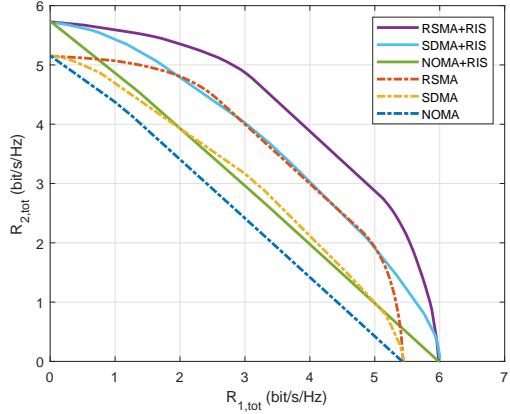}
	\caption{ IR rate region comparison of different strategies with $P_t=10 \text{dB}$, $E_{th}=20\mu\text{W}$}
  \vspace{-4mm}
\end{figure}
\begin{figure}[t!]	
	\centering
	\includegraphics[width=0.43\textwidth]{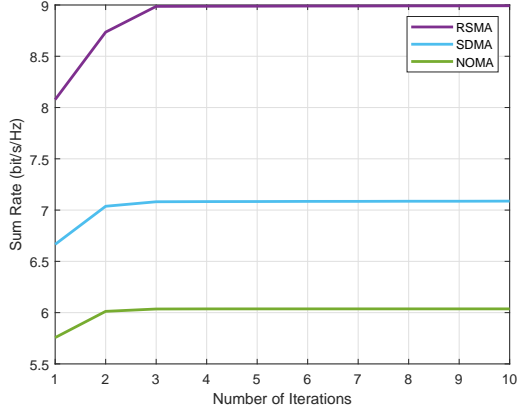}
	\caption{Convergence of the algorithms in one channel realization  with $P_t=10 \text{dB}$, $E_{th}=20\mu\text{W}$}
  \vspace{-4mm}
\end{figure}
\par 
RSMA+RIS, SDMA+RIS and NOMA+RIS can be solved by Algorithm 3, while RSMA, SDMA and NOMA can be solved directly using Algorithms 1.
\par In the simulation, the location distribution of BS, RIS, IRs
and ERs are shown in Fig. 2. To be specific, ERs are randomly distributed in a circle of radius $r_1=0.1$ meters and centered at (1, 0), IRs are randomly distributed in a circle of radius $r_1=1$ meter and centered at (20, 0). The path loss of all channels is set to $P(d) = Ld^{-\alpha}$, where $d$ represents the
distance between the BS and the IRs/ERs, $\alpha$ is the
path loss exponents. $L$ represents the signal attenuation at a
distance of 1 meter, which is usually set to -30 dB. In addition, the path loss exponents of BS to all IRs, BS to all ERs, BS
to RIS, RIS to all IRs, RIS to all ERs are set to 2, 3, 3,
3.5 and 1.5. Besides, the number of transmit antennas at the BS is $N_t=2$, and the number of IRs, ERs, and RIS elements is $K=J=2$, $N=8$, respectively. The energy conversion efficiency is $\zeta=0.5$, the noise of user $k$ is $\sigma^2_k = -80$ dBm and the convergence tolerance is $\epsilon =10^{-3}$. All simulation results
are averaged over 100 random channel realizations.
\par Fig. 3 shows the rate-energy tradeoff when $P_t=10$ dB and the weights of IRs are set to $u_1 = u_2 = 1$. It can be observed that when the collected energy $\text{E}_{th}=20\mu\text{W}$, the respective gain of RSMA+RIS over SDMA, NOMA, RSMA, SDMA+RIS and NOMA+RIS are 27.55\%, 38.81\%, 12.78\% 12.02\% and 27.69\%. The simulation results show that RSMA+RIS achieves explicit performance gain over SDMA, NOMA, RSMA, SDMA+RIS and  NOMA+RIS especially when the harvested energy constraint at the ERs is high. It achieves a better tradeoff between the WSR of IRs and the harvested energy at the ERs. 
\par Fig. 4 shows the rate regions of IRs for different strategies, with the assistance of RIS, the rate regions of all strategies are enlarged. One interesting observation is, the rate region of RSMA almost coincides with that of SDMA+RIS especially when the two users have similar weights. It further shows the promising capability of RSMA to reduce the transmission complexity while guaranteeing the quality of service.
\par Fig. 5 demonstrates the convergence of the proposed AO algorithm with $P_t=10\text{dB}$ and $E_{th}=20\mu\text{W}$. This graph shows that the proposed algorithm converges very quickly.
\section{Conclusion}
In this paper, we propose a SWIPT system empowered by RSMA and RIS. The precoders and RIS phase shifts are jointly designed at the transmitter to maximize the WSR of IRs under the harvested energy constraint of ERs. We propose an AO algorithm to solve this problem, which alternately optimizes the phase shift matrix and the remaining optimization variables. Numerical results shows that the proposed approach significantly improves the WSR and harvested energy tradeoff between the IRs and ERs. 
RIS empowered RSMA is therefore a promising strategy for SWIPT.
\bibliographystyle{IEEEtran}
\bibliography{reference}
\end{document}